\documentclass[acmsmall,screen]{acmart}
\usepackage[ruled,vlined]{algorithm2e} 
\usepackage{graphicx}
\usepackage{subcaption} 
\usepackage{amsmath}
\usepackage{multirow}
\usepackage{tabularx} 
\usepackage{makecell} 
\usepackage{enumitem}
\usepackage{color}
\usepackage{xcolor}
\usepackage{colortbl}
\usepackage{tcolorbox} 
\definecolor{mygray}{gray}{.9}

\NewDocumentCommand{\framecolorbox}{oommm}
 {
  \IfValueTF{#1}
   {\IfValueTF{#2}
    {\fcolorbox{#3}{#4}{\makebox[#1][#2]{#5}}}
    {\fcolorbox{#3}{#4}{\makebox[#1]{#5}}}%
   }
   {\fcolorbox{#3}{#4}{#5}}%
 }

\newtcolorbox{promptbox}[2][]{
    enhanced,
    before skip=10pt,
    after skip=10pt,
    colback=gray!5,        
    colframe=black!70,      
    boxrule=0.5pt,
    arc=1pt,                
    left=10pt, right=10pt, top=10pt, bottom=10pt,
    fonttitle=\bfseries\sffamily,
    coltitle=white,
    colbacktitle=black!70,
    title={#2},             
    breakable,              
    attach boxed title to top left={yshift=-2mm, xshift=2mm},
    boxed title style={sharp corners, size=small},
    #1                      
}

\acmSubmissionID{146} 
\acmDOI{10.1145/nnnnnnn.nnnnnnn}
\acmYear{2026} 
\setcopyright{acmlicensed}

\newcommand{\Rmnum}[1]{\uppercase\expandafter{\romannumeral #1}}

\newcommand{\Comment}[1]{}
\newcolumntype{P}{S[table-format=2.2]}
\newcolumntype{N}{S[table-format=4.0]}

\newcommand{\tech}{\textit{\normalsize{R}\footnotesize{E}\normalsize{CRL}}}

\author{Shouyu Yin}
\affiliation{%
  \institution{Tianjin University}
  \city{Tianjin}
  \country{China}
}
\email{yinshouyu@tju.edu.cn}

\author{Zhao Tian}
\affiliation{%
  \institution{Tianjin University}
  \city{Tianjin}
  \country{China}
}
\email{tianzhao@tju.edu.cn}

\author{Junjie Chen}
\authornote{Corresponding author.}
\affiliation{%
  \institution{Tianjin University}
  \city{Tianjin}
  \country{China}
}
\email{junjiechen@tju.edu.cn}

\author{Shikai Guo}
\affiliation{%
  \institution{Dalian Maritime University}
  \city{Dalian}
  \country{China}
}
\email{shikai.guo@dlmu.edu.cn}

\title{Improving LLM Code Generation via Requirement-Aware Curriculum Reinforcement Learning}

\begin{document}

\begin{abstract}
Code generation, which aims to automatically generate source code from given programming requirements, has the potential to substantially improve software development efficiency. 
With the rapid advancement of large language models (LLMs), LLM-based code generation has attracted widespread attention from both academia and industry. 
However, as programming requirements become increasingly complex, existing LLMs still exhibit notable performance limitations. 
To address this challenge, recent studies have proposed training-based curriculum reinforcement learning (CRL) strategies to improve LLM code generation performance. 
Despite their effectiveness, existing CRL approaches suffer from several limitations, including misaligned requirement difficulty perception, the absence of requirement difficulty optimization, and suboptimal curriculum sampling strategies.
In CRL-based code generation, programming requirements serve as the sole input to the model, making their quality and difficulty critical to training effectiveness. 
Motivated by insights from software requirements engineering, we propose \tech{}, a novel requirement-aware curriculum reinforcement learning framework for enhancing LLM-based code generation. 
\tech{} automatically perceives model-specific requirement difficulty, optimizes challenging requirements to improve training data utilization, and employs an adaptive curriculum sampling strategy to construct training batches with smoothly varying difficulty. 
Extensive experiments on five state-of-the-art LLMs across five widely-used code generation benchmarks by comparing with five state-of-the-art baselines, demonstrate the significant effectiveness of \tech{}.
For example, \tech{} achieves an average Pass@1 improvement of 1.23\%$\sim$5.62\% over all state-of-the-art baselines. 
\end{abstract}
 
\maketitle

\keywords{Code Generation, Large Language Models, Reinforcement Learning, Requirements Engineering}

\section{Introduction}
\label{sec:introduction}
Code generation refers to the automatic generation of source code from given programming requirements. 
Due to its potential to substantially improve software development efficiency and productivity, code generation has attracted increasing attention from both academia and industry in recent years~\cite{guo2024deepseekcode, chen2021humaneval, hendrycks2021MBPP}. 
With the rapid advancement of large language models (LLMs), such as Claude~\cite{anthropic2024claude3} and DeepSeek~\cite{liu2024deepseek}, LLM-based code generation has triggered a paradigm shift in intelligent software engineering~\cite{jiang2024survey, hou2024large}. 
Despite these advances, LLMs still exhibit notable performance limitations when confronted with complex programming requirements~\cite{hendrycks2021APPS}.
Such limitations not only hinder the practical deployment of LLM-based code generation techniques but may also lead to critical software quality issues~\cite{liu2023your, pearce2025asleep}. 
Consequently, effectively improving the code generation performance of LLMs remains a pressing research challenge.

Existing approaches to improving LLM-based code generation performance can be broadly divided into inference-time and training-time methods~\cite{gao2025trae, zhu2025uncertainty, dou2024stepcoder,guo2024deepseekcode}. 
Inference-time approaches (e.g., prompting engineering~\cite{wei2022chain} and agents~\cite{gao2025trae}) enhance code generation performance by refining the inference process; however, they are inherently constrained by the capabilities of the underlying base model~\cite{jiang2024survey}. 
To achieve more fundamental and sustained performance improvements, recent research has increasingly focused on training-time optimization strategies~\cite{ouyang2022training, le2022coderl}. 
Among these, curriculum reinforcement learning (CRL) has emerged as one of the most influential training paradigms for LLMs~\cite{narvekar2020curriculum, nair2024curriculum}. 
Originally proposed by ~\citet{bengio2009curriculum}, CRL organizes training tasks according to pre-defined difficulty levels and trains models progressively from easier to more challenging tasks, thereby improving learning efficiency and final performance.

Recently, CRL has been widely adopted to fundamentally enhance the performance of LLM-based code generation~\cite{parashar2025curriculum, nair2024curriculum}. 
In this context, each training instance corresponds to a programming requirement, and the model is trained following a curriculum schedule defined by task difficulty. 
Typically, an LLM receives a positive reward only when the generated code successfully passes all golden test cases. 
For example, ~\citet{parashar2025curriculum} explore CRL strategies based on manually-annotated difficulty labels and systematically evaluate different curriculum scheduling policies. 
~\citet{nair2024curriculum} apply CRL by labeling reference implementations using cyclomatic complexity~\cite{bengio2009curriculum} and Halstead difficulty~\cite{hendrycks2021measuring}, and systematically investigate the effects of multiple curriculum scheduling strategies on improving model capabilities.

However, existing CRL-based approaches overlook the perception and optimization of programming requirements, which serve as the sole model input during training and fundamentally determine both training efficiency and generation performance. 
First, existing CRL approaches rely on fixed requirement difficulty labels, either manually annotated or inferred from static code complexity metrics, which fail to capture the dynamic, model-specific perception of requirement difficulty. 
This results in a misalignment between the pre-defined requirement difficulty and the actual requirement difficulty perceived by different models. 
Second, existing CRL approaches do not adequately address highly challenging requirements that exceed the model's current capability. 
For such challenging requirements, all the corresponding sampled codes tend to be incorrect, yielding no positive reward signals and substantially reducing training data utilization~\cite{hendrycks2021APPS}.
Third, existing CRL sampling strategies often introduce abrupt transitions in task difficulty, which can cause catastrophic forgetting of previously-learned knowledge and induce cold-start issues at new curriculum stages, ultimately degrading training stability and effectiveness. 
These limitations highlight the need for a more principled approach to requirement difficulty perception, optimization, and adaptive curriculum sampling in CRL-based code generation.

To address these challenges, we propose the first requirement-aware curriculum reinforcement learning framework for enhancing LLM-based code generation, called \textbf{\tech{}} (\textbf{RE}quirement-aware \textbf{C}urriculum \textbf{R}einforcement \textbf{L}earning). 
It improves the performance of LLM code generation by automatically perceiving and optimizing requirement difficulty and adaptively constructing training batches that align with different curriculum stages. 
\tech{} consists of three key components: \textit{requirement difficulty perception}, \textit{requirement difficulty optimization}, and \textit{adaptive curriculum sampling}.
First, to dynamically obtain accurate model-perceived requirement difficulty, \tech{} integrates code generation with automated testing, using execution correctness to quantify the difficulty of a given requirement for the base model. 
Second, to improve the utilization of challenging requirements, \tech{} introduces requirement optimization and revision agents that optimize requirement difficulty based on five pre-defined software requirement attributes, ensuring that these challenging requirements can yield more informative reward signals during training. 
Third, to mitigate catastrophic forgetting and cold-start effects, \tech{} employs an adaptive curriculum sampling strategy with a difficulty smoothing mechanism, which gradually exposes the model to increasingly challenging requirements while maintaining rehearsal of previously learned knowledge. 
Together, these components enhance the CRL process and substantially improve LLM performance in code generation. 
Owing to its modular and extensible design, \tech{} also holds promise for broader applications in other complex software engineering tasks.

We comprehensively evaluate \tech{} on five state-of-the-art open-source LLMs (i.e., Qwen2.5-Coder-1.5B, Qwen2.5-Coder-3B, Qwen2.5-Coder-7B, Llama-3.2-3B, and SmolLM3-3B) using five widely-adopted code generation benchmarks (i.e., HumanEval, HumanEval+, MBPP, MBPP+, and LiveCodeBench). 
Experimental results demonstrate that \tech{} consistently outperforms five state-of-the-art baselines across all 25 subjects (5 LLMs $\times$ 5 benchmarks), validating the effectiveness of requirement-aware CRL in improving LLM-based code generation. 
For instance, \tech{} achieves an average Pass@1 improvement of 1.23\%$\sim$5.62\% over all baselines across all 25 subjects. 
In particular, \tech{} improves the average Pass@1 of the smaller Qwen2.5-Coder-3B from 60.07\% to 66.50\%, enabling it to slightly outperform the substantially larger Qwen2.5-Coder-7B (66.05\%).
Moreover, we further conduct comprehensive ablation studies using four variants of \tech{}, confirming the necessity and effectiveness of each main component. 
Additionally, we analyze the impact of a key hyper-parameter (i.e., the difficulty smoothing factor) on training performance, demonstrating the robustness and rationality of our default setting.

The main contributions of this paper are summarized as follows:
\begin{itemize}[leftmargin=10pt]
    \item \textbf{Novel Perspective}: 
    We introduce a novel perspective for enhancing LLM-based code generation through the sophisticated requirement-aware curriculum reinforcement learning framework.
    
    \item \textbf{Tool Implementation}: 
    We implement \tech{} following the novel perspective, which integrates model-perceived requirement difficulty estimation, requirement difficulty optimization, and adaptive curriculum sampling to improve training stability and effectiveness.
        
    \item \textbf{Extensive Evaluation}: 
    We conduct large-scale experiments on five advanced open-source LLMs and five widely-used benchmarks, demonstrating that \tech{} consistently outperforms five state-of-the-art baselines and significantly improves code generation performance.
\end{itemize}

\section{Background and Related Work}
\label{sec:background}

\subsection{LLM-based Code Generation}

Code generation aims to automatically synthesize executable programs that satisfy a given programming requirement, typically expressed in natural language~\cite{yin2017syntactic}.
Recent advances in LLMs have fundamentally reshaped the code generation paradigm~\cite{chen2021humaneval, roziere2023code}. 
Formally, \emph{LLM-based code generation} can be modeled as a conditional sequence generation problem~\cite{sutskever2014sequence}.
Given a requirement $r$ specified in natural language (optionally augmented with contextual information such as partial code or documentation), an LLM parameterized by $\theta$ generates a code sequence $c = (c_1, \ldots, c_T)$ by maximizing the conditional likelihood $p_\theta(c \mid r)$.
In recent years, open-source LLMs (e.g., Qwen-Coder~\cite{hui2024qwen2}, Llama~\cite{grattafiori2024llama3}, and SmolLM~\cite{bakouch2025smollm3}) leverage large-scale pre-training, long-context modeling, and instruction alignment to substantially improve code generation performance.
These LLMs effectively bridge the gap between high-level requirements and executable code.

Despite their popularity, LLMs still exhibit notable performance limitations, particularly when handling complex programming requirements~\cite{liu2023evalplus}. 
To enhance the code generation capabilities of LLMs, a number of techniques have been proposed in recent years. 
In general, these techniques can be categorized into two complementary paradigms: inference-time and training-time methods.
Inference-time techniques improve LLM performance in code generation on-the-fly by introducing prompt-based and agent-based strategies. 
For example, ~\citet{zhu2025uncertainty} propose UnCert-CoT, an uncertainty-aware Chain-of-Thought method that enhances code generation by explicitly incorporating model uncertainty into the reasoning process. 
~\citet{zhang2023repocoder} introduce RepoCoder, which streamlines repository-level code completion by integrating a similarity-based retriever in an iterative retrieval-augmented generation pipeline. 
In addition, ~\citet{hong2023metagpt} develope MetaGPT, a multi-agent framework that decomposes the code generation task into collaborative sub-tasks executed by specialized agents organized in a structured pipeline. 
More recently, ~\citet{gao2025trae} propose Trae Agent, an LLM-based agent for general-purpose software engineering tasks, which can interpret natural language instructions and autonomously execute complex workflows by invoking external tools.

On the other hand, training-time techniques enhance code generation performance by improving the quality of training data or designing effective training strategies (e.g., supervised fine-tuning and reinforcement learning). 
For example, ~\citet{huang2025opencoder} propose a meticulous data cleaning strategy to extract high-quality code from large-scale web corpora, thereby improving training effectiveness. 
~\citet{luo2024wizardcoder} introduce Evol-Instruction, which employs an automated evolutionary process to generate high-quality fine-tuning datasets with varying levels of complexity, enabling more effective adaptation of code LLMs.
In the context of reinforcement learning, ~\citet{shojaee2023execution} propose PPOCoder, which applies Proximal Policy Optimization (PPO)~\cite{schulman2017proximal} combined with execution feedback to ensure stable policy updates and improve functional correctness. 
In particular, our study focuses on curriculum reinforcement learning, a widely-adopted paradigm that arranges training data from easy to difficult, gradually adapting the model to increasingly complex programming requirements and thereby promoting an effective and stable training process.

\subsection{Curriculum Reinforcement Learning for LLM Code Generation}

Curriculum reinforcement learning (CRL), originally introduced by Bengio et al.~\cite{bengio2009curriculum}, improves reinforcement learning by progressively organizing training tasks from easy to difficult, thereby facilitating more effective model training. 
In essence, CRL formulates learning as an adaptive task-sequencing problem. 
Given a task space (${x_i} \subset \mathcal{X}$), the policy model $\pi_\theta$ does not sample training instances uniformly. 
Instead, training follows a curriculum schedule $\mathcal{C}$ that governs the exposure to tasks according to an increasing difficulty order. 
For each sampled task $x_i$, the model generates an output $y_i \sim \pi_\theta(\cdot \mid x_i)$, and receives a learning signal from a reward function $R(x_i, y_i)$, which quantifies prediction quality with respect to task-specific constraints. 
CRL has become a widely-adopted and effective training strategy for LLMs, demonstrating strong empirical performance across diverse tasks~\cite{qi2024webrl, nair2024curriculum, xi2024training}.

In recent years, CRL has been widely applied to code generation tasks~\cite{dou2024stepcoder, nair2024curriculum}. 
Specifically, each task is instantiated as a \textit{programming requirement} $r_i$~\cite{hendrycks2021APPS, hendrycks2021MBPP}. 
Each requirement is associated with a difficulty level, which can be annotated by human experts~\cite{puri2021codenet} or inferred from the code complexity of the corresponding reference implementation~\cite{nair2024curriculum}. 
These requirements are then organized into a curriculum schedule $\mathcal{C}$, enabling a step-wise learning process that gradually exposes the model to increasingly complex programming scenarios. 
During training, LLMs typically receive positive reward signals only when the generated code successfully passes all golden tests, making execution correctness the primary learning objective. 
Under this paradigm, the quality and granularity of programming requirements play a crucial role in shaping generation performance, as well-specified requirements significantly increase the likelihood of producing functionally correct code~\cite{austin2021program, ma2025specgen}.

Numerous CRL techniques have been proposed to improve LLM performance in code generation~\cite{dou2024stepcoder, nair2024curriculum, parashar2025curriculum, sagtani2025improving}.
For example, ~\citet{dou2024stepcoder} propose StepCoder, which decomposes long-horizon code generation tasks into a sequence of code completion sub-tasks, thereby constructing an effective curriculum gradient for CRL strategy.
~\citet{parashar2025curriculum} explore CRL strategies using human-annotated difficulty labels, providing a comprehensive evaluation of different curriculum scheduling policies and their impact on training effectiveness. 
~\citet{nair2024curriculum} apply CRL by labeling reference code using cyclomatic complexity~\cite{bengio2009curriculum} and Halstead difficulty~\cite{hendrycks2021measuring}, and systematically investigate the effects of multiple curriculum scheduling strategies on improving model capabilities.
Despite these advances, existing CRL approaches suffer from several limitations in code generation task. 
First, requirement difficulty perception is often misaligned with the evolving capacity of the model, as it relies on static complexity metrics. 
Second, existing methods overlook challenging requirements that the model currently fails to resolve, resulting in suboptimal training data utilization. 
Furthermore, CRL training frequently encounters catastrophic forgetting and cold-start effects, which undermine training stability and slow convergence. 
These limitations prevent LLMs from fully exploiting the potential of available code data.

\begin{figure}[t]
    \centering
    \begin{subfigure}{0.32\textwidth}
        \centering
        \includegraphics[width=\linewidth]{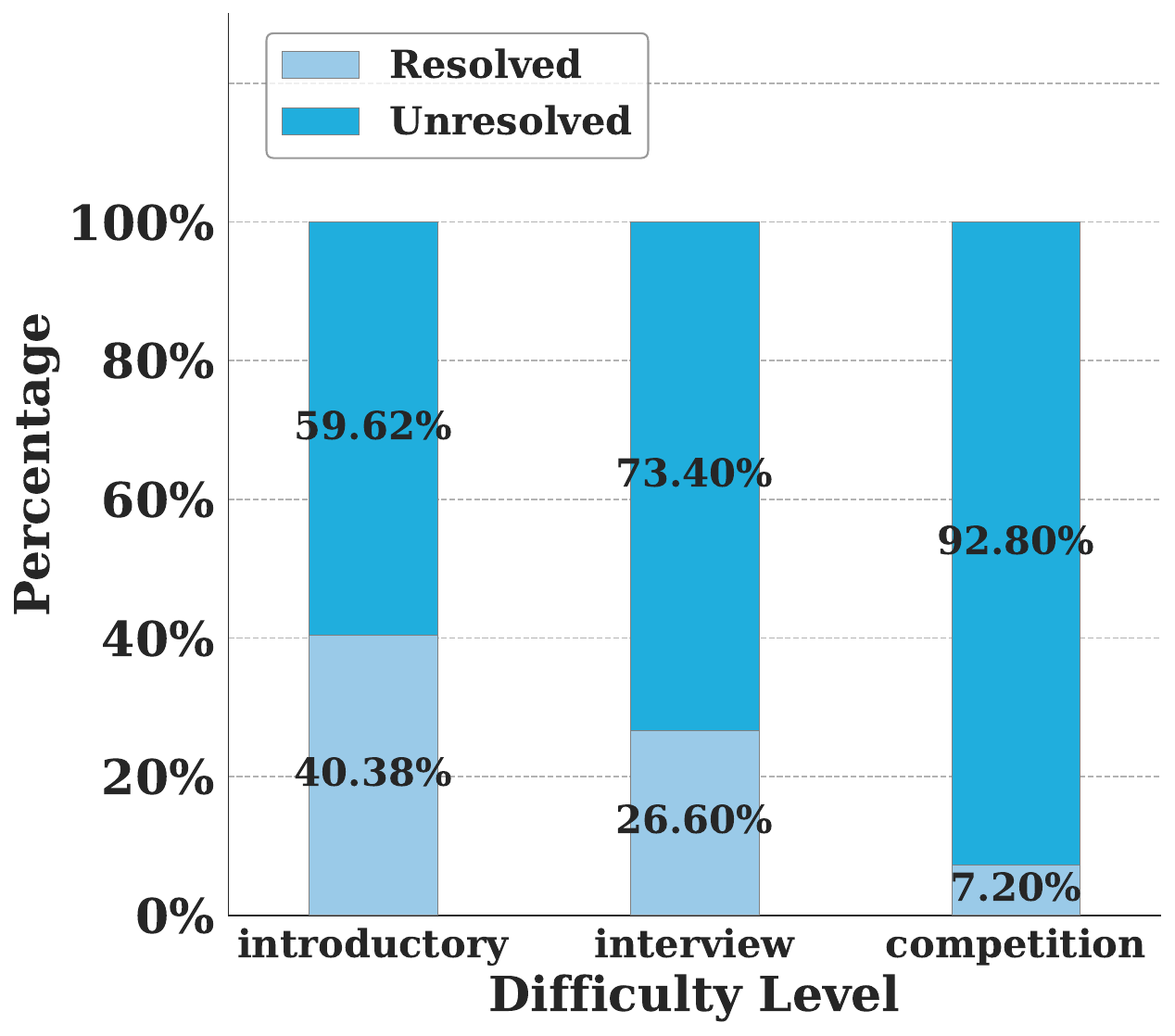}
        \caption{ }
        \label{fig:motivation:sub_a}
    \end{subfigure}
    \hfill
    \begin{subfigure}{0.32\textwidth}
        \centering
        \includegraphics[width=\linewidth]{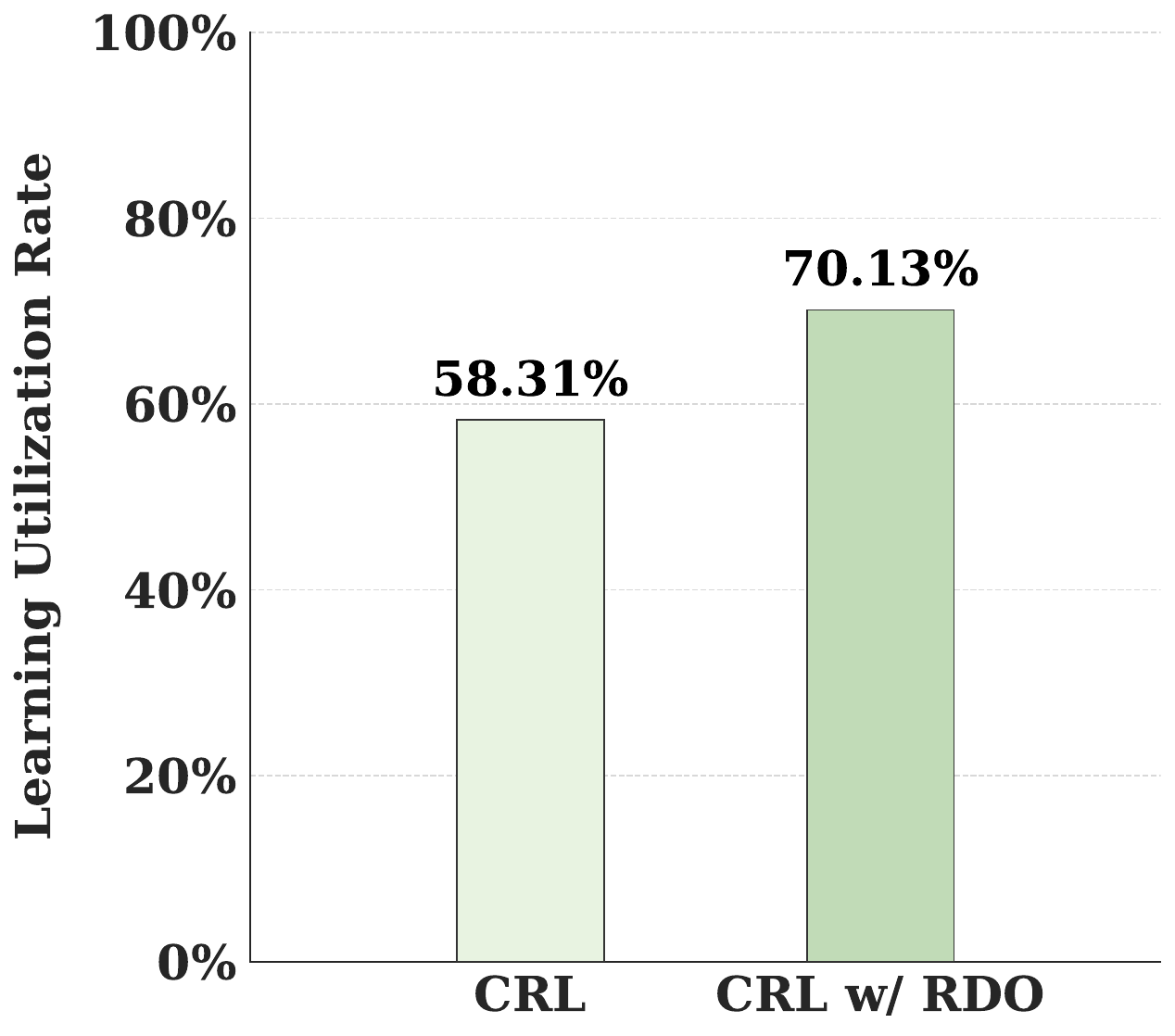}
        \caption{ }
        \label{fig:motivation:sub_b}
    \end{subfigure}
    \hfill
    \begin{subfigure}{0.32\textwidth}
        \centering
        \includegraphics[width=\linewidth]{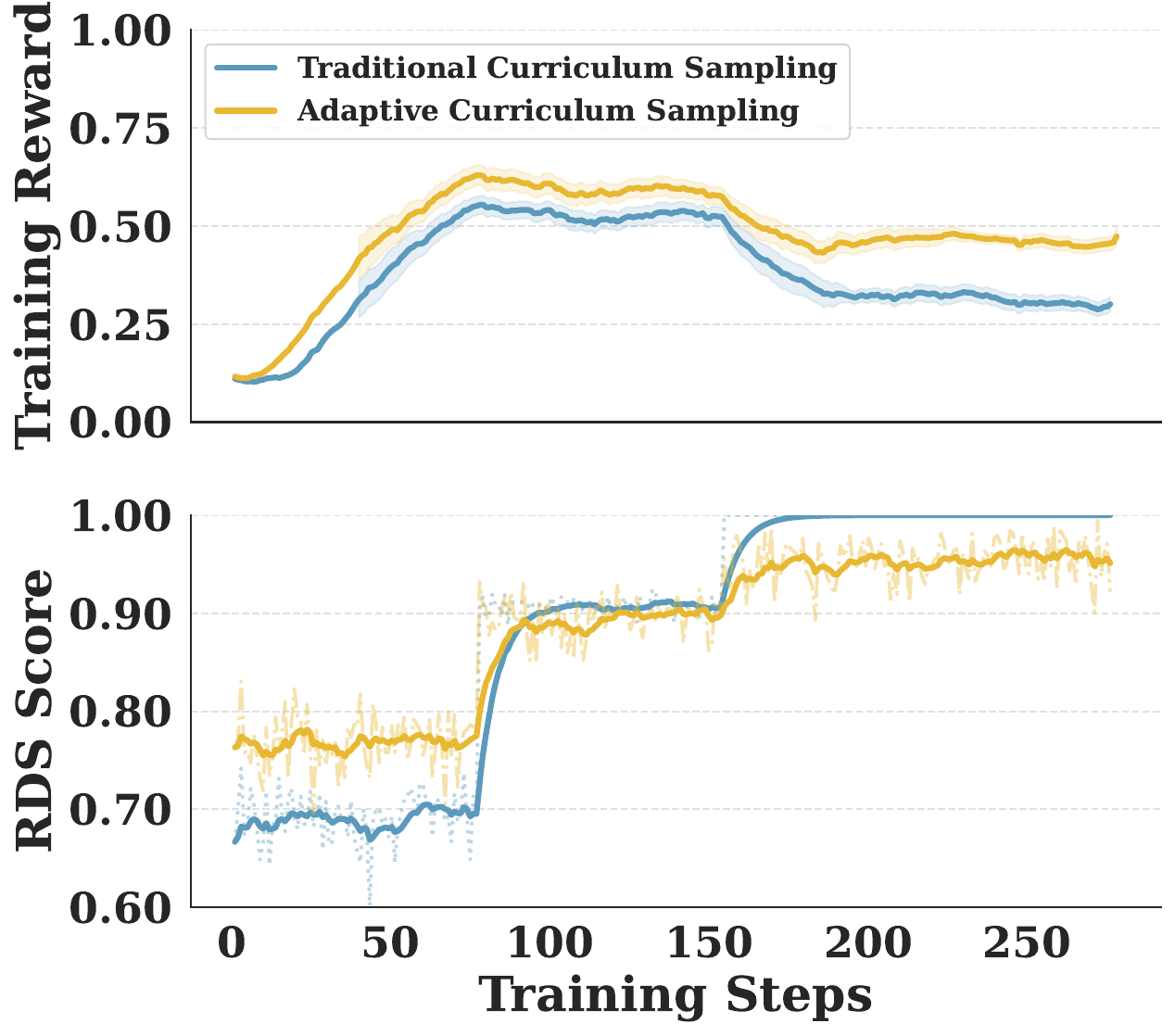}
        \caption{ }
        \label{fig:motivation:sub_c}
    \end{subfigure}
    \caption{The preliminary experiments on the APPS+ benchmark using Qwen2.5-Coder-3B: 
    (a) discrepancy of manually-annotated difficulty and model-perceived difficulty; 
    (b) contribution of requirement difficulty  optimization in terms of learning utilization rate; 
    and (c) effectiveness comparison between traditional curriculum sampling and adaptive curriculum sampling.}
    \label{fig:three_plots}
\end{figure}

\section{Motivation}
\label{sec:motivation}

To motivate the necessity of our three key components (requirement difficulty perception, requirement difficulty optimization, and adaptive curriculum sampling), we conduct a set of preliminary experiments on the APPS+ dataset using Qwen2.5-Coder-3B, with results summarized in Figure~\ref{fig:three_plots}.

First, existing difficulty perception approaches typically rely on manually-annotated difficulty scores. 
To examine the alignment between such annotations and model-perceived difficulty, we analyze their discrepancy as shown in Figure~\ref{fig:motivation:sub_a}. 
Specifically, we perform 16 independent sampling runs for each problem using Qwen2.5-Coder-3B on APPS+. 
An instance is labeled as ``resolved'' if at least one sampled solution passes all golden tests; otherwise, it is labeled as ``unresolved''. 
APPS+ categorizes programming requirements into three manually-annotated difficulty levels (i.e., ``introductory'', ``interview'', and ``competition'').
Our analysis reveals a substantial misalignment between manual annotations and model-perceived difficulty. 
Even for the lowest difficulty level (i.e., ``introductory''), the model fails to solve up to 59.62\% of the instances (i.e., produces incorrect solutions in all 16 samples). 
This result indicates that manually-annotated difficulty scores do not accurately reflect the actual model-perceived requirement difficulty. 
Moreover, such static difficulty annotations cannot adapt to models with different capacities or training stages. 
\textit{This motivates the necessity of our requirement difficulty perception component, which dynamically and accurately perceives the model-specific requirement difficulty.}

Second, the challenge arises from the presence of a large proportion of programming requirements that exceed the model's current capability. 
As shown in Figure~\ref{fig:motivation:sub_a}, fewer than 7.2\% of challenging requirements (i.e., ``competition'') can be successfully solved. 
These extremely difficult instances consistently yield zero reward signals (i.e., no successful codes across 16 samples), leading to a low \textit{learning utilization rate}, defined as the proportion of training instances for which at least one sampled solution achieves non-zero correctness. 
Such underutilized data significantly impairs training effectiveness.
Figure~\ref{fig:motivation:sub_b} shows that the learning utilization rate of the original APPS+ is only 58.31\%. 
After introducing our requirement difficulty optimization component, this rate increases to 70.13\%, indicating more effective use of challenging requirements. 
\textit{This motivates the necessity of our requirement difficulty optimization component, which substantially improves training data utilization and overall training performance.}

Finally, we examine the limitations of traditional curriculum sampling strategies. 
As illustrated in Figure~\ref{fig:motivation:sub_c}, traditional curriculum sampling often introduces abrupt shifts in task difficulty, as measured by the RDS metric defined in Section~\ref{sec:DP_component}. 
Such sharp difficulty transitions can lead to catastrophic forgetting of previously learned knowledge and exacerbate the cold-start problem when the model is exposed to a new difficulty stage during training.
By contrast, as shown in Figure~\ref{fig:motivation:sub_c}, our adaptive curriculum sampling strategy produces a significantly smoother RDS trajectory and effectively mitigates the sharp reward drops commonly observed in traditional curriculum sampling. 
This smoother progression stabilizes training and facilitates more consistent performance improvements. 
\textit{This motivates the necessity of our adaptive curriculum sampling component, which dynamically adjusts the sampling distribution to improve training stability and effectiveness.}

\section{Approach}
\label{sec:approach}


\subsection{Overview}

\begin{figure*}[t]
    \caption{The overview of \tech{}}
    \label{fig:method}
    \centering
    \includegraphics[width=\textwidth]{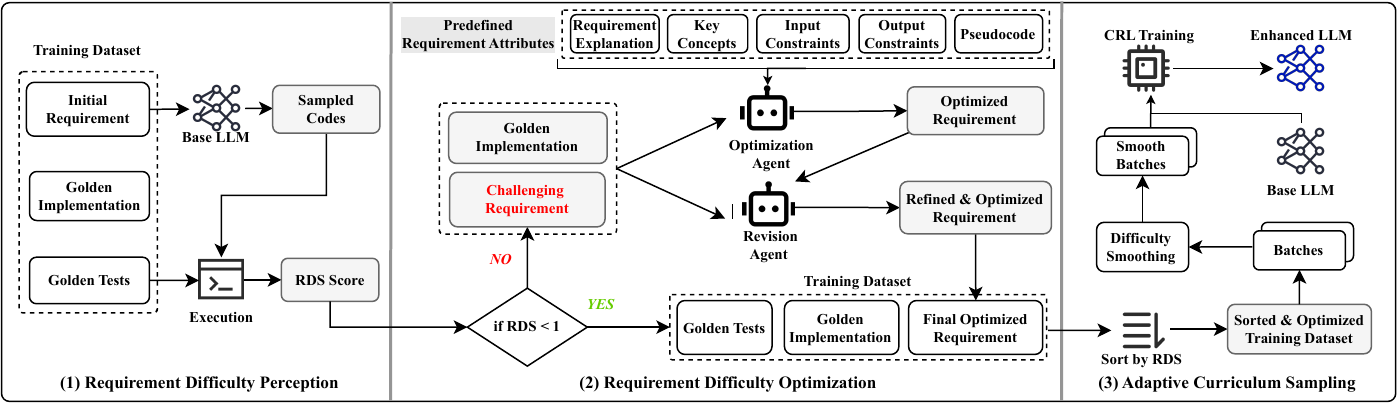}
\end{figure*}

In this paper, we propose \tech{}, a requirement-aware CRL technique designed to substantially improve LLM performance in code generation.
Specifically, \tech{} automatically obtains LLM-perceived requirement difficulty, optimizes challenging requirements, and designs the adaptive curriculum sampling strategy to construct stage-aware training batches, leading to improved code generation performance. 
An overview of \tech{} is illustrated in Figure~\ref{fig:method}.
\tech{} consists of three key components:
(1) \textbf{Requirement Difficulty Perception} (Section~\ref{sec:DP_component}) combines code generation and automated testing to assess the functional correctness of programs generated from initial requirements to determine the LLM-perceived difficulties for programming requirements;
(2) \textbf{Requirement Difficulty Optimization} (Section~\ref{subsec:RDO_component}) designs the requirement optimization and revision agents that further refines challenging requirements, thereby reducing their LLM-perceived difficulties to obtain more stable and positive reward signals during CRL training;
(3) \textbf{Adaptive Curriculum Sampling} (Section~\ref{sec:HS_component}) dynamically constructs training batches that align with different stages of CRL based on model-perceived requirement difficulty, while introducing a difficulty smoothing mechanism to mitigate catastrophic forgetting and encourage gradual exposure to harder requirements.
Based on these three components, \tech{} effectively enhances the CRL process and significantly improves LLM performance in code generation. 
In the following sections, we provide a detailed description of each component in \tech{}.



\subsection{Requirement Difficulty Perception}
\label{sec:DP_component}

In CRL framework, the accurate perception of requirement difficulty is a critical step, as it directly influences the effectiveness of subsequent curriculum sampling and training, and ultimately affects model performance~\cite{narvekar2020curriculum}.
However, existing CRL methods often suffer from inadequate difficulty perception, primarily due to a misalignment between pre-defined difficulty and model-perceived difficulty.
First, mainstream manually-annotated difficulty labels~\cite{puri2021codenet, hendrycks2021measuring} typically rely on expert experience and domain knowledge; therefore, they fail to accurately reflect the actual requirement difficulty perceived by models.
Similarly, code complexity–based difficulty scores~\cite{nair2024curriculum}, which are derived from software quality metrics such as code size or runtime complexity, do not reflect the model-perceived requirement difficulty.
As discussed in Section~\ref{sec:motivation}, a significant number of programming tasks (59.62\%) that are manually labeled as ``introductory'' or associated with low code complexity cannot be easily solved by LLMs.
Moreover, both types of difficulty scores are invariant, whereas the difficulty of the same programming requirement can vary substantially across models with different architectures, parameter scales, and capabilities.

As illustrated in Figure~\ref{fig:method}, we propose a model-specific requirement difficulty perception strategy that leverages the model's code generation performance on training tasks to construct difficulty scores.
Specifically, for each programming requirement, we first use the base model to independently sample $N$ candidate codes.
We then execute the corresponding golden tests in the training dataset to verify the functional correctness of each sampled code.
The average code correctness rate serves as an indicator of how difficult the requirement is for the model.
Formally, we define the \textbf{Requirement Difficulty Score} (RDS) as:
\begin{equation}
    \label{eq:rds}
    RDS = 1 - \frac{1}{N} \sum_{i=i}^N Correctness(code_i)
\end{equation}
where $code_i$ denotes the $i$-th sampled program ($1 \leq i \leq N$) generated by the target model for the given programming requirement, and $Correctness(code_i)$ is a binary indicator that equals 1 if the program passes all golden tests and 0 otherwise.
The resulting RDS value lies in the range $[0, 1]$, where larger values indicate higher requirement difficulty for the model.

Unlike conventional difficulty metrics, RDS is directly grounded in model capacity, as it is computed from the correctness of model-generated codes based on the initial requirement.
Notably, the training reward in CRL is also derived from code correctness~\cite{le2022coderl}.
Therefore, RDS provides a semantically aligned and intuitive measure of requirement difficulty that is consistent with the CRL objective.
Particularly, even for the same programming requirement, different models may yield different RDS values, as the correctness of generated code varies with model capability.
This property makes RDS universally applicable across different models.

\subsection{Requirement Difficulty Optimization}
\label{subsec:RDO_component}


After perceiving the model-specific requirement difficulty, we analyze the requirement distribution over the training set (in Table~\ref{fig:motivation:sub_a}) and observe that a substantial fraction (i.e., 69.52\%) of requirements have an RDS of 1.0.
This indicates that all $N$ codes sampled by the base model for these \textit{challenging requirements} are incorrect. 
Following existing studies~\cite{narvekar2020curriculum}, these challenging requirements pose a major obstacle for CRL. 
Since all sampled codes are incorrect, they yield no positive reward signals, preventing the model from effectively learning from these instances and substantially reducing training effectiveness. 
To address this issue, we introduce the requirement difficulty optimization component that aims to improve training data utilization by effectively reducing the difficulty of challenging requirements, thereby enhancing overall training effectiveness.

Inspired by principles from software requirements engineering~\cite{sommerville2011software}, we design a novel \textbf{requirement optimization agent} (as shown in Figure~\ref{fig:method}) to optimize the difficulty of challenging requirements.
Specifically, the requirement optimization agent takes as input the original programming requirement and its corresponding golden code implementation, and outputs an optimized requirement.
Inspired by established practices in requirements specification~\cite{sommerville2011software}, we define five structured requirement attributes, including (1) \textit{requirement explanation}, (2) \textit{key concepts}, (3) \textit{input constraints}, (4) \textit{output constraints}, and (5) \textit{pseudocode}.
By explicitly incorporating these attributes, the requirement optimization agent produces standardized and information-enriched requirements that expose critical implementation details from the golden code, thereby reducing the model-perceived difficulty of the original requirements and improving its ability to generate correct code solutions.
We further discuss the scalability of these requirement attributes in Section~\ref{subsec:future_work}, highlighting the potential to enhance the generality and extensibility of our approach.

To further mitigate potential hallucinations introduced by the requirement optimization agent, we design a \textbf{requirement revision agent} that validates and refines the initial optimized requirements.
Specifically, the requirement revision agent takes as input the original programming requirement, the initial optimized requirement, and the golden implementation.
It checks for ambiguity, redundancy, logical inconsistencies, and incompleteness, and produces a final optimized requirement.
After obtaining the optimized requirements for challenging requirements, we recompute their RDS values.
Empirically, this component substantially increases the proportion of challenging requirements that the model can successfully solve by 35.25\%, confirming that this component enhances practical performance significantly.
For cases where the difficulty remains unchanged, we retain the original requirements to avoid introducing misleading or erroneous information.

\subsection{Adaptive Curriculum Sampling}
\label{sec:HS_component}

In traditional CRL~\cite{narvekar2020curriculum}, training data are always partitioned into multiple clusters according to difficulty (e.g., from easy to hard), with each cluster corresponding to a distinct training stage. 
During each stage, the model samples exclusively from the cluster associated with the current difficulty level, and transitions to the next stage upon completion. 
While intuitive, such curriculum sampling strategies suffer from two key limitations. 
First, stage transitions are often abrupt, causing a cold-start problem~\cite{narvekar2020curriculum} when the model encounters substantially harder data in a new difficulty stage, which degrades training stability and effectiveness.
Second, after incorporating new knowledge at higher difficulty levels, the model tends to forget previously learned knowledge from earlier stages, leading to catastrophic forgetting~\cite{aleixo2024catastrophic}.

To address these challenges, we propose an adaptive curriculum sampling component (as shown in Figure~\ref{fig:method}). 
The core idea is to dynamically construct training batches that are aligned with different stages of CRL based on model-perceived requirement difficulty, and introduce a \textbf{requirement difficulty smoothing mechanism} to mitigate catastrophic forgetting and encourage gradual exposure to harder requirements.
Specifically, we first sort all training instances by their RDS values, and partition them into three curriculum clusters following prior work~\cite{nair2024curriculum}. 
We then employ the difficulty smoothing mechanism, which samples data not only from the current-stage cluster but also draws a small proportion of samples from the other clusters. 
This hybrid strategy allows the model to both resample previously-learned easier requirements (supporting memory replay) and encounter more challenging requirements in advance, thereby alleviating cold-start effects when transitioning between stages.

To precisely control this difficulty smoothing mechanism, we introduce a difficulty smoothing factor $\lambda \in [0, 1]$, which governs the curriculum sampling ratio. 
At a given training stage, a proportion $\lambda$ of each batch is sampled from the current difficulty cluster, while the remaining proportion $(1-\lambda)$ is sampled uniformly from the entire training set. 
Larger values of $\lambda$ encourage the model to focus more heavily on the current curriculum stage, whereas smaller values increase curriculum smoothness by amplifying cross-stage sampling.
The difficulty smoothing factor $\lambda$ is a key hyper-parameter of \tech{}, and its impact is analyzed in Section~\ref{subsec:rq3}. 
Through this adaptive curriculum sampling strategy, \tech{} achieves more stable and effective training dynamics, thereby improving overall model performance.

\section{Evaluation Design}
\label{sec:evaluation}

Our study aims to address the following three research questions (RQs):
\begin{itemize}[leftmargin=10pt]
    \item \textbf{RQ1}: How does \tech{} perform in terms of effectiveness compared to state-of-the-art techniques?
    \item \textbf{RQ2}: How does each main component in \tech{} contribute to the overall effectiveness?
    \item \textbf{RQ3}: How does the difficulty smoothing factor (i.e., $\lambda$) and sampling temperature (i.e., $T$) affect \tech{}'s effectiveness?
\end{itemize}

\subsection{Studied LLMs}
To comprehensively evaluate \tech{}, we select five state-of-the-art open-source LLMs that cover three representative model architectures and three different parameter scales.
Specifically, to examine the effectiveness of \tech{} across different model architectures, we include three 3B-parameter models: Qwen2.5-Coder-3B~\cite{hui2024qwen2}, Llama-3.2-3B~\cite{grattafiori2024llama3}, and SmolLM3-3B~\cite{bakouch2025smollm3}.
These models represent three mainstream, trainable open-source LLMs and provide strong code generation performance at a moderate and widely-used parameter scale.
In addition, to analyze the impact of model scale, we further evaluate three sizes from the Qwen2.5-Coder family, including Qwen2.5-Coder-1.5B, Qwen2.5-Coder-3B, and Qwen2.5-Coder-7B.
The selection of these model sizes is primarily motivated by the practical consideration of training cost,as well as their widespread adoption in prior research~\cite{guo2024deepseekcode, nair2024curriculum, chen2025self}.
Overall, all selected models are widely used in code generation tasks and collectively represent state-of-the-art open-source LLMs~\cite{hui2024qwen2, grattafiori2024llama3, bakouch2025smollm3}.


%

\subsection{Benchmarks}
\label{subsec:benchmarks}
To comprehensively evaluate \tech{}, we utilize five widely-recognized benchmarks in our study: 
HumanEval~\cite{chen2021humaneval}, HumanEval+~\cite{liu2023evalplus}, MBPP~\cite{hendrycks2021MBPP}, MBPP+~\cite{liu2023evalplus}, and LiveCodeBench~\cite{jain2024livecodebench}. 
These benchmarks have been extensively adopted in existing studies to evaluate the performance of LLMs in code generation~\cite{dou2024stepcoder, le2022coderl, wei2025swe_rl, wang2025codeboost}.
Subsequently, we provide a detailed description of each benchmark.

\textbf{HumanEval} contains 164 human-written programming problems proposed by OpenAI.
Each instance consists of a function signature, a natural language requirement, and several golden tests for automated correctness validation (averaging 7.7 golden tests per instance).
We also use the extended version, \textbf{HumanEval+}, which includes an average of 754 additional tests over the original version.
More rigorous tests could determine the correctness of generated code more precisely.

\textbf{MBPP} comprises 974 crowd-sourced Python programming tasks proposed by Google Research, each with natural language requirements and three golden tests.
Besides, we use \textbf{MBPP+}, an extended version of MBPP, by fixing ambiguous descriptions and complementing an average of 104 tests for each programming task.
Following the existing work~\cite{liu2023evalplus}, we use all 378 programming tasks from the test sets of both MBPP and MBPP+.

\textbf{LiveCodeBench} contains 1,055 challenging programming problems sourced from competitive programming platforms (i.e., LeetCode, AtCoder, and Codeforces).
It dynamically reflects the LLM performance in more complex algorithmic reasoning, intricate I/O requirements, and stricter execution constraints. 
In our experiments, we use all 1,055 programming problems.

\textbf{APPS+} is an extended version of APPS~\cite{hendrycks2021APPS} (which is collected from various competitive programming platforms), meticulously excluding instances with code syntax errors and low-quality descriptions.
Specifically, APPS+ consists of 7,456 programming tasks comprising three difficulty levels (i.e., introductory, interview, and competition). 

Following existing work~\cite{dou2024stepcoder}, we adopt APPS+ to support CRL for each studied LLM by splitting it into 80\% for training and 20\% for validation, and then evaluate the updated models exclusively on the remaining five benchmarks.
This experimental design is motivated by three considerations.
First, APPS+ contains a substantially larger number of instances than the other five benchmarks combined, enabling statistically meaningful training and validation splits. 
Second, reserving all other benchmarks for the testing sets effectively prevents data leakage, since they are never used during model training and validation.
Third, APPS+ provides manually annotated difficulty labels, which we adopt as a baseline (see Section~\ref{subsec:baselines}) to systematically compare human-constructed difficulty labels with difficulty labels automatically inferred by \tech{}, thereby enabling a controlled assessment of the effectiveness of our requirement difficulty perception component.

\subsection{Metrics}

We execute the golden tests (provided by benchmarks) to assess the functional correctness of the code generated for each programming problem, and report Pass@$k$ and AvgPassRatio as our primary evaluation metrics. 
Both metrics are widely adopted in existing studies and provide complementary evaluation of LLM performance in code generation~\cite{jiang2024self,tian2025fixing,tian2025aligning}

\textbf{Pass@$k$}~\cite{chen2021humaneval} evaluates whether a programming problem is solved by at least one of the $k$ generated solutions. 
Specifically, for each problem, the LLM produces $k$ candidate solutions, and the problem is considered solved if any candidate passes all evaluation test cases. 
Pass@$k$ is then defined as the percentage of solved problems over the total number of problems in the entire benchmark. 
Following the existing work~\cite{chen2021humaneval}, developers tend to consider and inspect one code solution produced by the LLM, and thus we report \textbf{Pass@1} as the primary setting. 
Note that Pass@1 is a more stringent criterion, making improvements in Pass@1 particularly challenging and meaningful. 
Larger Pass@$k$ values indicate better code generation correctness.

\textbf{AvgPassRatio}~\cite{hao2022aixbench} captures the degree of code correctness rather than binary success or failure. 
For each programming problem, it computes the fraction of test cases passed by the generated solution, and then averages this ratio across all problems. 
While Pass@$k$ reflects whether a problem is fully solved, AvgPassRatio quantifies partial correctness, making the two metrics complementary. 
Larger AvgPassRatio values indicate better code generation performance.

\subsection{Compared Techniques}
\label{subsec:baselines}

To comprehensively evaluate \tech{}, we select three representative or state-of-the-art \textbf{CRL baselines} that vary in difficulty perception and curriculum sampling strategies.


\begin{itemize}[leftmargin=10pt]
    \item \textbf{Random Sampling-based Curriculum Reinforcement Learning (RSCRL)}~\cite{bengio2009curriculum}:
    It randomly samples each training batch from the entire training dataset at every iteration, without considering any difficulty annotations associated with the programming tasks.
    
    \item \textbf{Manual Difficulty-based Curriculum Reinforcement Learning (MDCRL)}~\cite{hendrycks2021measuring}:
    It leverages the manually-annotated difficulty labels provided in the APPS+ dataset to perform staged reinforcement learning, progressively training the model from simpler to more difficult tasks.
    
    \item \textbf{Overall Metric-based Curriculum Reinforcement Learning (OMCRL)}~\cite{nair2024curriculum}: 
    It assesses task difficulty using a composite metric defined as the average of cyclomatic complexity~\cite{mccabe1976complexity} and Halstead difficulty~\cite{halstead1977elements}. 
    During training, it resamples instances from previous curriculum stages to construct a difficulty-mixed curriculum schedule.
\end{itemize}


In addition, as the requirement difficulty optimization component of \tech{} employs more advanced models as auxiliary agents, we further include two \textbf{knowledge distillation–based baselines}. 
This experimental design allows us to disentangle the performance gains attributable to the requirement optimization mechanism itself from those arising solely from distilled knowledge, thereby ensuring a fair and comprehensive evaluation.
\begin{itemize}[leftmargin=10pt]
    \item \textbf{COTTON}~\cite{yang2024COTTON}: 
    It adopts a multi-agent chain-of-thought alignment framework to distill high-quality reasoning trajectories from GPT-3.5-Turbo through systematic quality and consistency checks. 
    Based on the distilled data, COTTON enables smaller models to achieve performance comparable to, or even surpassing, that of larger LLMs.
    
    \item \textbf{CodePLAN}~\cite{sun2024CODEPLAN}:
    It distills structured solution plans from a teacher model by jointly leveraging programming requirements and reference implementations. 
    It employs a multi-task training objective that simultaneously optimizes code generation and plan prediction, allowing smaller models to substantially outperform conventional supervised fine-tuning baselines.    
\end{itemize}


In summary, RSCRL, MDCRL, and OMCRL facilitate a systematic comparison between \tech{} and representative CRL approaches, with OMCRL serving as a state-of-the-art CRL baseline.
COTTON and CodePLAN enable a complementary comparison with knowledge distillation–based methods, among which COTTON represents the current state of the art.

\subsection{Implementation Details}
During the training phase, we adopt the VeRL framework~\cite{sheng2025verl}, a widely-used RL framework for LLMs. 
Specifically, each training instance is sampled five times to estimate the advantage. 
The number of training epochs, batch size, maximum response length, decoding temperature, and learning rate are set to 2, 24, 1024, 1.0, and $3 \times 10^{-5}$, respectively. 
Following prior work~\cite{hu2022lora}, we employ Low-Rank Adaptation (LoRA) to reduce GPU memory consumption during training.
Both the LoRA decomposition rank ($r$) and scaling factor ($\alpha$) are set to 32, which is a commonly used configuration that balances micro-adjustment capacity with memory efficiency for medium-sized models. 
For the requirement difficulty perception component, we set the sampling size $N=16$ to balance computational efficiency with estimation accuracy.
For the adaptive curriculum sampling component, we set the difficulty smoothing factor $\lambda=0.6$, which was determined through experiments in Section~\ref{subsec:RQ3}.
During the evaluation phase, to eliminate randomness and ensure reproducibility, we fix the decoding temperature to 0. 
Due to space limitations, the detailed prompt design of agents in \tech{} can be found in our anonymous repository~\cite{rercl_code}.

\section{Results and Analysis}
\label{sec:results}


\begin{table}[htbp]
  \centering
  \tabcolsep=1.5mm
  \caption{Effectiveness comparison in terms of Pass@1 ($\uparrow$).}
  \label{tab:rq1_pass_results}
  \small
  \renewcommand{\arraystretch}{1.1}
  \begin{tabular}{llccccccc}
    \toprule
    \textbf{LLM} & \textbf{Technique} & \makecell{\textbf{HumanEval}} & \makecell{\textbf{HumanEval+}} & \textbf{MBPP} & \makecell{\textbf{MBPP+}} & \makecell{\textbf{LCB}} & \textbf{Avg.} & \makecell{\textbf{$\Delta$ ($\uparrow$)}} \\
    \midrule
    \multirow{7}{*}{Qwen-1.5B} & Base & 68.90\% & 62.20\% & 72.75\% & 61.90\% & 2.84\% & 53.72\% & - \\
    \cmidrule{2-9}
    & CodePlan & 69.51\% & 63.41\% & 72.22\% & 62.70\% & 8.34\% & 55.24\% & 1.52\% \\
    & COTTON & 70.73\% & 64.02\% & 72.75\% & 62.96\% & 9.19\% & 55.93\% & 2.21\%\\
    \cmidrule{2-9}
    & RSCRL & 70.73\% & 64.02\% & 73.81\% & 62.70\% & 11.47\% & 56.55\% & 2.83\% \\
    & MDCRL & 71.34\% & 65.24\% & 73.54\% & 62.43\% & 10.62\% & 56.63\% & 2.92\% \\
    & OMCRL & 70.73\% & 64.63\% & \textbf{75.66\%} & 64.02\% & 10.24\% & 57.06\% & 3.34\% \\
    & \tech{} & \textbf{74.39\%} & \textbf{67.07\%} & 74.07\% & \textbf{64.55\%} & \textbf{11.94\%} & \textbf{58.40\%} & \textbf{4.68\%} \\
    \midrule 
    \multirow{7}{*}{Qwen-3B} & Base & 75.61\% & 66.46\% & 76.46\% & 65.61\% & 16.21\% & 60.07\% & - \\
    \cmidrule{2-9}
    & CodePlan & 74.39\% & 67.07\% & 76.72\% & 66.40\% & 19.81\% & 60.88\% & 0.81\% \\
    & COTTON & 79.27\% & 70.73\% & 75.93\% & 66.14\% & 20.66\% & 62.55\% & 2.48\% \\
    \cmidrule{2-9}
    & RSCRL & 81.10\% & 72.56\% & 75.40\% & 65.08\% & 22.75\% & 63.38\% & 3.31\% \\
    & MDCRL & 82.32\% & 73.78\% & 75.13\% & 65.87\% & 23.70\% & 64.16\% & 4.09\% \\
    & OMCRL & 81.71\% & 73.17\% & \textbf{79.63\%} & \textbf{67.46\%} & 24.36\% & 65.27\% & 5.20\% \\
    & \tech{} & \textbf{82.93\%} & \textbf{77.44\%} & 79.10\% & \textbf{67.46\%} & \textbf{25.59\%} & \textbf{66.50\%} & \textbf{6.43\%} \\
    \midrule
    \multirow{7}{*}{Qwen-7B}  & Base & 83.54\% & 77.44\% & 80.42\% & 69.05\% & 19.81\% & 66.05\% & - \\
    \cmidrule{2-9}
    & CodePlan & 84.15\% & 78.05\% & 80.95\% & 70.11\% & 21.61\% & 66.97\% & 0.92\% \\
    & COTTON & 84.76\% & 79.88\% & 81.48\% & 70.37\% & 21.52\% & 67.60\% & 1.55\% \\
    \cmidrule{2-9}
    & RSCRL & 85.37\% & 78.05\% & 82.54\% & 70.37\% & 26.92\% & 68.65\% & 2.60\%  \\
    & MDCRL & 86.59\% & 79.27\% & 82.80\% & 70.37\% & 27.77\% & 69.36\% & 3.31\% \\
    & OMCRL & \textbf{87.80\%} & 81.10\% & 80.95\% & 69.58\% & 28.34\% & 69.55\% & 3.50\% \\
    & \tech{} & 87.20\% & \textbf{82.32\%} & \textbf{83.86\%} & \textbf{70.63\%} & \textbf{31.28\%} & \textbf{71.06\%} & \textbf{5.01\%} \\ 
    \midrule
    \multirow{7}{*}{Llama3.2-3B} & Base & 54.27\% & 48.78\% & 61.11\% & 50.26\% & 10.24\% & 44.93\% & - \\
    \cmidrule{2-9}
    & CodePlan & 60.37\% & 53.05\% & 62.96\% & 52.91\% & 12.32\% & 48.32\% & 3.39\% \\
    & COTTON & 59.76\% & 54.88\% & 63.49\% & 53.97\% & 12.70\% & 48.96\% & 4.03\% \\
    \cmidrule{2-9}
    & RSCRL & 59.15\% & 52.44\% & 65.34\% & 54.23\% & 11.66\% & 48.56\% & 3.63\% \\
    & MDCRL & 61.59\% & \textbf{56.10\%} & 63.49\% & 53.44\% & 12.51\% & 49.43\% & 4.50\% \\
    & OMCRL & 62.20\% & 54.27\% & 65.87\% & 55.56\% & 13.27\% & 50.23\% & 5.30\% \\
    & \tech{} & \textbf{63.41\%} & 55.49\% & \textbf{68.78\%} & \textbf{57.14\%} & \textbf{14.88\%} & \textbf{51.94\%} & \textbf{7.01\%} \\
    \midrule
    \multirow{7}{*}{SmoLlm-3B} & Base & 67.07\% & 59.15\% & 62.70\% & 53.44\% & 14.50\% & 51.37\% & - \\
    \cmidrule{2-9}
    & CodePlan & 67.68\% & 61.59\% & 63.49\% & 53.44\% & 18.01\% & 52.84\% & 1.47\% \\
    & COTTON & 65.85\% & 58.54\% & 62.96\% & 54.23\% & 19.43\% & 52.20\% & 0.83\% \\
    \cmidrule{2-9}
    & RSCRL & 67.68\% & 64.02\% & 61.64\% & 52.12\% & 20.95\% & 53.28\% & 1.91\% \\
    & MDCRL & \textbf{70.12\%} & 64.63\% & 64.29\% & 55.03\% & 21.14\% & 55.04\% & 3.67\% \\
    & OMCRL & 68.29\% & 64.02\% & 63.76\% & 53.97\% & 22.75\% & 54.56\% & 3.19\% \\
    & \tech{} & 68.90\% & \textbf{65.85\%} & \textbf{65.61\%} & \textbf{55.82\%} & \textbf{22.94\%} & \textbf{55.82\%} & \textbf{4.45\%} \\
    \bottomrule
  \end{tabular}
\end{table}

\begin{table}[htbp]
  \centering
  \tabcolsep=1.5mm
  \caption{Effectiveness comparison in terms of AvgPassRatio ($\uparrow$).}
  \label{tab:rq1_avg_results}
  \small
  \renewcommand{\arraystretch}{1.1}
  \begin{tabular}{llccccccc}
    \toprule
    \textbf{LLM} & \textbf{Technique} & \makecell{\textbf{HumanEval}} & \makecell{\textbf{HumanEval+}} & \textbf{MBPP} & \makecell{\textbf{MBPP+}} & \makecell{\textbf{LCB}} & \textbf{Avg.} & \makecell{\textbf{$\Delta$ ($\uparrow$)}} \\
    \midrule
    \multirow{7}{*}{Qwen-1.5B} & Base & 79.66\% & 81.01\% & 75.20\% & 74.96\% & 13.34\% & 64.83\% & - \\
    \cmidrule{2-9}
    & CodePlan &79.41\% &82.15\% &74.12\% &75.88\% &19.50\% &66.21\% &1.38\% \\
    & COTTON &80.25\% &81.90\% &75.45\% &76.10\% &20.15\% &66.77\% &1.94\% \\
    \cmidrule{2-9}
      & RSCRL &80.80\% &82.05\% &76.90\% &75.30\% &24.80\% &67.97\% &3.14\% \\
      & MDCRL &82.90\% &82.50\% &76.80\% &75.10\% &23.90\% &68.24\% &3.41\% \\
      & OMCRL &83.35\% &84.80\% &\textbf{77.95\%} &77.20\% &24.45\% &69.55\% &4.72\% \\
      & \tech{} &\textbf{84.65\%} &\textbf{87.71\%} &77.56\% &\textbf{78.32\%} &\textbf{26.10\%} &\textbf{70.87\%} &\textbf{6.03\%} \\
    \midrule
    \multirow{7}{*}{Qwen-3B} & Base &84.51\% &86.15\% &76.46\% &78.29\% &26.90\% &70.46\% &- \\
    \cmidrule{2-9}
      & CodePlan &82.90\% &87.10\% &79.20\% &79.05\% &29.85\% &71.62\% &1.16\% \\
      & COTTON &87.10\% &89.55\% &76.60\% &79.40\% &33.10\% &73.15\% &2.69\% \\
    \cmidrule{2-9}
      & RSCRL &88.95\% &91.40\% &76.85\% &80.10\% &33.20\% &74.10\% &3.64\% \\
      & MDCRL &91.30\% &92.80\% &77.10\% &80.95\% &34.15\% &75.26\% &4.80\% \\
      & OMCRL &90.85\% &92.90\% &81.94\% &81.20\% &36.29\% &76.64\% &6.17\% \\
      & \tech{} &\textbf{91.77\%} &\textbf{93.64\%} &\textbf{82.25\%} &\textbf{82.39\%} &\textbf{38.99\%} &\textbf{77.81\%} &\textbf{7.35\%} \\
    \midrule
    \multirow{7}{*}{Qwen-7B} & Base &90.02\% &90.82\% &82.45\% &81.98\% &28.23\% &74.70\% &- \\
    \cmidrule{2-9}
      & CodePlan &91.15\% &90.45\% &83.50\% &82.80\% &31.50\% &75.88\% &1.18\% \\
      & COTTON &90.90\% &93.10\% &83.10\% &83.05\% &32.40\% &76.51\% &1.81\% \\
    \cmidrule{2-9}
      & RSCRL &92.45\% &91.05\% &84.10\% &83.25\% &35.80\% &77.33\% &2.63\% \\
      & MDCRL &92.66\% &92.70\% &84.60\% &83.15\% &36.40\% &77.90\% &3.20\% \\
      & OMCRL &\textbf{93.05\%} &94.20\% &84.55\% &\textbf{83.95\%} &39.15\% &78.98\% &4.28\% \\
      & \tech{} &93.02\% &\textbf{95.55\%} &\textbf{85.01\%} &83.86\% &\textbf{41.40\%} &\textbf{79.77\%} &\textbf{5.07\%} \\
    \midrule
    \multirow{7}{*}{Llama3.2-3B} & Base &67.60\% &69.74\% &64.73\% &65.38\% &20.40\% &57.57\% &- \\
    \cmidrule{2-9}
      & CodePlan &74.80\% &74.20\% &67.10\% &67.50\% &22.85\% &61.29\% &3.72\% \\
      & COTTON &75.10\% &77.90\% &66.50\% &69.85\% &24.10\% &62.69\% &5.12\% \\
    \cmidrule{2-9}
      & RSCRL &73.15\% &75.30\% &68.20\% &71.10\% &21.65\% &61.88\% &4.31\% \\
      & MDCRL &77.40\% &\textbf{80.20\%} &66.85\% &68.90\% &22.40\% &63.15\% &5.58\% \\
      & OMCRL &76.90\% &78.15\% &68.45\% &73.50\% &25.30\% &64.46\% &6.89\% \\
      & \tech{} &\textbf{78.99\%} &79.33\% &\textbf{72.83\%} &\textbf{74.33\%} &\textbf{26.50\%} &\textbf{66.40\%} &\textbf{8.83\%} \\
    \midrule
    \multirow{7}{*}{SmoLlm-3B} & Base &79.13\% &80.50\% &64.18\% &64.61\% &18.01\% &61.29\% &- \\
    \cmidrule{2-9}
      & CodePlan &79.55\% &82.40\% &65.30\% &64.30\% &22.45\% &62.80\% &1.51\% \\
      & COTTON &78.05\% &79.90\% &64.85\% &65.80\% &26.35\% &62.99\% &1.70\% \\
    \cmidrule{2-9}
      & RSCRL &80.20\% &86.55\% &62.80\% &63.45\% &27.90\% &64.18\% &2.89\% \\
      & MDCRL &\textbf{82.53\%} &87.10\% &66.45\% &65.95\% &33.85\% &67.18\% &5.89\% \\
      & OMCRL &81.10\% &86.80\% &64.95\% &66.15\% &38.88\% &67.58\% &6.29\% \\
      & \tech{} &81.43\% &\textbf{89.80\%} &\textbf{66.94\%} &\textbf{67.49\%} &\textbf{43.64\%} &\textbf{69.86\%} &\textbf{8.57\%} \\
    \bottomrule
  \end{tabular}
\end{table}


\subsection{RQ1: Overall Effectiveness of \tech{}}

\subsubsection{Process:}
To answer RQ1, we apply \tech{} and five baselines (i,e., RSCRL, MDCRL, OMCRL, COTTON, and CodePlan) to each of the five studied LLMs (i.e., Qwen2.5-Coder-1.5B, Qwen2.5-Coder-3B, Qwen2.5-Coder-7B, Llama-3.2-3B, and SmolLM3-3B).
All LLMs are fine-tuned using APPS+, with the pre-defined training and validation splits, and subsequently evaluated on the remaining five benchmarks (i.e., HumanEval, HumanEval+, MBPP, MBPP+, and LiveCodeBench) used exclusively for testing. 
The effectiveness of each technique is measured using two code correctness metrics (i.e., Pass@1 and AvgPassRatio). 
To ensure a fair comparison, all techniques are evaluated under the same hyper-parameter settings, with the decoding temperature fixed to 0 and the number of training epochs is set to 2 for CRL methods and 5 for knowledge distillation-based methods. 
For \tech{}'s requirement difficulty optimization component and both knowledge distillation-based baselines, we use DeepSeek-V3 to build the agent or serve as the teacher model.

\subsubsection{Results:}
Tables~\ref{tab:rq1_pass_results} and ~\ref{tab:rq1_avg_results} report the comparison results of all studied techniques in terms of Pass@1 and AvgPassRatio, respectively.
We first observe that, on average, CRL-based methods consistently outperform knowledge distillation–based approaches on both metrics,demonstrating the effectiveness of CRL for LLMs in code generation. 
In particular, it indicates that the performance gains of \tech{} are not primarily attributed to the distilled reasoning signals from teacher models.


Additionally, \tech{} achieves the best average performance across all studied LLMs. 
On average, \tech{} improves over all baselines by 1.23\%$\sim$5.62\% in Pass@1 and 0.79\%$\sim$7.06\% in AvgPassRatio, demonstrating its effectiveness and generalizability.
Meanwhile, the few cases in which \tech{} exhibit limited or degraded performance primarily occur on HumanEval and MBPP. 
These benchmarks are relatively simple, and the base models already achieve strong performance, leaving limited headroom for further improvement. 
As a result, performance tends to approach saturation, making the training more sensitive to stochastic variations and leading to minor fluctuations.


We further observe a consistent trend that \tech{} substantially narrows the performance gap between small and large models. 
Across all studied LLMs, smaller models benefit more from the CRL process, achieving relative improvements that bring their performance closer to that of larger LLMs. 
For instance, \tech{} improves the average Pass@1 of Qwen2.5-Coder-3B from 60.07\% to 66.50\%, slightly surpassing the original performance of the much larger Qwen2.5-Coder-7B (66.05\%).  
And \tech{} improves the average Pass@1 of Qwen2.5-Coder-1.5B from 53.72\% to 58.40\%, narrowing the gap with Qwen2.5-Coder-3B (60.07\%) to only 1.67\%. 
Moreover, on the HumanEval+ benchmark, Qwen2.5-Coder-1.5B achieves a Pass@1 of 67.07\%, outperforming Qwen2.5-Coder-3B, which attains 66.46\%.
This systematic enhancement indicates that \tech{} can significantly boost the capability of smaller models, enabling them to approach or even exceed the performance of larger models.
We plan to further investigate and validate this scalability advantage on larger LLMs in future work.

Moreover, among the CRL baselines, MDCRL and OMCRL outperform RSCRL on average.
Specifically, MDCRL and OMCRL improve Pass@1 by an average of 0.84\% and 1.25\%, and AvgPassRatio by 1.25\% and 2.35\%, respectively, compared to RSCRL. 
These results highlight the importance of effective difficulty perception strategies for CRL methods. 
Meanwhile, MDCRL performs worse than RSCRL on MBPP and MBPP+, underperforming by 1.85\% and 0.79\% on Llama3.2-3B, respectively.
This degradation likely stems from the misalignment of manually-annotated difficulty labels.
Notably, \tech{} consistently outperforms both MDCRL and OMCRL, further demonstrating the advantage of our designed requirement-aware strategies for CRL methods.

Across different model scales, \tech{} consistently improves performance, with the largest relative gains seen in small and medium scales. 
For instance, on Qwen2.5-Coder-1.5B, \tech{} increases the average Pass@1 by 4.68\%, while on Qwen2.5-Coder-3B, it raises to 66.50\%, slightly surpassing the performance of original Qwen2.5-Coder-7B. 
Therefore, CRL is especially effective for smaller models, helping them narrow the performance gap with larger models while providing steady gains across all scales.

\begin{table}[t]
  \centering
  \tabcolsep=1.7mm
  \caption{Comparison between \tech{} and its variants in terms of Pass@1 ($\uparrow$) and AvgPassRatio ($\uparrow$).}
  \label{tab:abaltion}
    \begin{tabular}{lcccccc}
    \toprule
    \textbf{Technique} & \textbf{HumanEval} & \textbf{HumanEval+}  & \textbf{MBPP}  & \textbf{MBPP+} & \textbf{LiveCodeBench} & \textbf{Avg.} \\
    \midrule
    \multicolumn{7}{c}{\framecolorbox[14cm][c]{gray!30}{gray!30}{\textbf{Pass@1 Metric}}} \\ \midrule
    Base & 75.61\% & 66.46\% & 76.46\% & 65.61\% & 16.21\% & 60.07\%  \\
    RSCRL & 81.10\% & 72.56\% & 75.40\% & 65.08\% & 22.75\% & 63.38\% \\
    \tech{}$_{wMA}$ & 80.49\% & 76.83\% & 78.31\% & 67.46\% & 23.32\% & 65.28\% \\
    \tech{}$_{wOM}$ & 82.32\% & 76.22\% & 79.10\% & 67.20\% & 24.93\% & 65.95\% \\
    \tech{}$_{woRDO}$ & 81.71\% & 75.61\% & 75.66\% & 66.40\% & 24.55\% & 64.79\% \\ 
    \tech{}$_{woRDS}$ & 79.88\% &76.22\% &75.40\% &65.34\% &24.64\% &64.30\% \\
    \tech{} & \textbf{82.93\%} & \textbf{77.44\%} & \textbf{79.10\%} & \textbf{67.46\%} & \textbf{25.59\%} & \textbf{66.50\%} \\
    \midrule
    \multicolumn{7}{c}{\framecolorbox[14cm][c]{gray!30}{gray!30}{\textbf{AvgPassRatio Metric}}} \\ \midrule
    Base & 84.51\% & 86.15\% & 76.46\% & 78.29\% & 26.90\% & 70.46\%  \\
    RSCRL & 88.95\% & 91.40\% & 76.85\% & 80.10\% & 33.20\% & 74.10\% \\
    \tech{}$_{wMA}$ & 89.92\% &92.50\% &81.35\% &81.60\% &35.50\% &76.17\% \\
    \tech{}$_{wOM}$ & 90.01\% & 91.23\% & \textbf{82.32\%} & 82.11\% & 37.73\% & 76.68\% \\
    \tech{}$_{woRDO}$ & 90.65\% &91.35\% &79.40\% &81.15\% &36.85\% &75.88\% \\
    \tech{}$_{woRDS}$ & 89.25\% &92.05\% &79.10\% &80.45\% &37.10\% &75.59\% \\
    \tech{} &\textbf{91.77\%} &\textbf{93.64\%} &82.25\% &\textbf{82.39\%} &\textbf{38.99\%} &\textbf{77.81\%} \\
    \bottomrule
    \end{tabular}%
  \label{tab:addlabel}%
\end{table}%


\subsection{RQ2: Contribution of Each Main Component in \tech{}}

\subsubsection{Variants:}
To comprehensively analyze the individual contributions of the three main components in \tech{} (i.e., requirement difficulty perception, requirement difficulty optimization, and adaptive curriculum sampling), we construct three ablated variants of \tech{}.
To assess the contribution of the requirement difficulty perception component, we replace the model-perceived requirement difficulty with manually-annotated difficulty labels from the training data and Overall-Metric labels from OMCRL, yielding \textbf{\tech{}$_{wMA}$} and \textbf{\tech{}$_{wOM}$}, respectively.
To evaluate the contribution of the requirement difficulty optimization component, we remove this component entirely, thereby disabling the difficulty optimization of challenging requirements and constructing \textbf{\tech{}$_{woRDO}$}.
To examine the contribution of the requirement difficulty smoothing mechanism in the adaptive curriculum sampling component, we remove it, resulting in \textbf{\tech{}$_{woRDS}$}.
For all three variants, we use the model with the maximum improvement determined in RQ1 (i.e., Qwen2.5-Coder-3B) as the base LLM and evaluate their performance on five benchmarks (i.e., HumanEval, HumanEval+, MBPP, MBPP+, and LiveCodeBench) in terms of both Pass@1 and AvgPassRatio.
All hyper-parameter settings are kept identical to those in RQ1 to ensure a fair comparison.

\subsubsection{Results:}
Table~\ref{tab:abaltion} reports the comparative performance of \tech{} and its four ablated variants in terms of Pass@1 and AvgPassRatio. 
We first observe that \tech{} consistently outperforms all three variants across all benchmarks. 
On average, \tech{} improves Pass@1 by 1.22\%, 0.55\%, 1.71\%, and 2.20\% over \tech{}$_{wMA}$, \tech{}$_{wOM}$,\tech{}$_{woRDO}$, and \tech{}$_{woRDS}$, respectively, and improves AvgPassRatio by 1.64\%, 1.13\%, 1.93\%, and 2.22\% correspondingly. 
These results demonstrate that each of the three components contributes positively to the overall effectiveness of \tech{}.

Additionally, all four variants still outperform the base model and RSCRL baseline on average. 
In particular, \tech{}$_{wMA}$, \tech{}$_{wOM}$, \tech{}$_{woRDO}$, and \tech{}$_{woRDS}$ achieve average improvements of 1.70\% in Pass@1 and 1.98\% in AvgPassRatio over the RSCRL. 
This observation further confirms that even partial instantiations of \tech{} retain performance benefits, while the full design yields the strongest gains.


Among the four variants, \tech{} consistently outperforms \tech{}$_{wMA}$ and \tech{}$_{wOM}$, demonstrating the stable effectiveness of our requirement difficulty perception component. 
Additionally, \tech{}$_{woRDO}$ shows a larger decrease, proving that optimizing requirements allows the model to better learn more challenging requirements, enhancing its code generation capabilities. 
In particular, \tech{}$_{woRDS}$ exhibits the largest relative decrease, demonstrating that it not only contributes independently but also interacts synergistically with requirement difficulty perception and requirement difficulty optimization, thereby enhancing their effectiveness.
Overall, these ablation results demonstrate that all main components are necessary for achieving the full performance gains of \tech{}, and that their joint integration is crucial to the significant improvement.

\begin{table}[t]
  \centering
  \tabcolsep=1.3mm
  \caption{Influence of the difficulty smoothing factor ($\lambda$) in terms of Pass@1 ($\uparrow$) and AvgPassRatio ($\uparrow$).}
  \label{tab:hyper_exp}
    \begin{tabular}{lcccccc}
    \toprule
    \textbf{Technique}  & \textbf{HumanEval} & \textbf{HumanEval+}  & \textbf{MBPP}  & \textbf{MBPP+} & \textbf{LiveCodeBench} & \textbf{Avg.} \\ \midrule
    \multicolumn{7}{c}{\framecolorbox[13.4cm][c]{gray!30}{gray!30}{\textbf{Pass@1 Metric}}} \\ \midrule
    \tech{}$_{\lambda=0}$ & 80.49\% & 76.22\% & 76.72\% & 64.81\% & 23.03\% & 64.25\% \\
    \tech{}$_{\lambda=0.2}$ & 81.10\% & 76.22\% & 77.78\% & 66.40\% & 23.98\% & 65.10\% \\
    \tech{}$_{\lambda=0.4}$ & 82.32\% & 75.61\% & 77.51\% & 66.14\% & 24.64\% & 65.24\% \\
    \tech{}$_{\lambda=0.6}$ & \textbf{82.93\%} & \textbf{77.44\% }& \textbf{79.10\%} & \textbf{67.46\%} & \textbf{25.59\%} & \textbf{66.50\%} \\
    \tech{}$_{\lambda=0.8}$ & 81.10\% & 73.78\% & 78.04\% & 65.08\% & 23.51\% & 64.30\% \\
    \tech{}$_{\lambda=1.0}$ & 79.88\% &76.22\% &75.40\% &65.34\% &24.64\% &64.30\%  \\ \midrule
    \multicolumn{7}{c}{\framecolorbox[13.4cm][c]{gray!30}{gray!30}{\textbf{AvgPassRatio Metric}}} \\ \midrule
    \tech{}$_{\lambda=0}$ & 89.15\% &91.80\% &79.86\% &79.53\% &34.20\% &74.91\% \\
    \tech{}$_{\lambda=0.2}$ & 90.10\% &92.09\% &80.95\% &81.38\% &35.82\% &76.07\% \\
    \tech{}$_{\lambda=0.4}$ & 91.20\% &91.17\% &80.62\% &80.92\% &37.47\% &76.28\% \\
    \tech{}$_{\lambda=0.6}$ & \textbf{91.77\%} &\textbf{93.64\%} &\textbf{82.25\%} &\textbf{82.39\%} &\textbf{38.99\%} &\textbf{77.81\%} \\
    \tech{}$_{\lambda=0.8}$ & 89.90\% &89.24\% &81.27\% &79.87\% &36.69\% &75.39\% \\
    \tech{}$_{\lambda=1.0}$ & 89.25\% &92.05\% &79.10\% &80.45\% &37.10\% &75.59\% \\
    \bottomrule
    \end{tabular}%
  \label{tab:addlabel}%
\end{table}%

\subsection{RQ3: Influence of the Difficulty Smoothing Factor and Sampling Temperature}
\label{subsec:rq3}

\subsubsection{Process:}
The difficulty smoothing factor is a key hyper-parameter of \tech{}, controlling the proportion of perceived difficulty-based sampled data relative to randomly sampled data within each training batch. 
Particularly, its formulation and role are described in detail in Section~\ref{sec:HS_component}.
Furthermore, the sampling temperature ($T$), as a critical hyperparameter during the inference phase, significantly influences the model's output. In practical applications, a temperature value greater than zero is typically selected to introduce diversity in the reasoning process.
In this research question, we investigate the influence of the difficulty smoothing factor (i.e., $\lambda$) and inference temperature on the effectiveness of \tech{}.
We consider six settings ranging from 0 to 1.0 with a step size of 0.2 (i.e., $\lambda \in \{0, 0.2, 0.4, 0.6, 0.8, 1.0\}$).
For each setting, we evaluate \tech{} on Qwen2.5-Coder-3B across all five benchmarks (i.e., HumanEval, HumanEval+, MBPP, MBPP+, LiveCodeBench), using Pass@1 and AvgPassRatio as evaluation metrics. 
All other experimental configurations are kept identical to those in RQ1 and RQ2 to ensure a controlled and fair comparison.

\subsubsection{Results:}
\label{subsec:RQ3}
Table~\ref{tab:hyper_exp} reports the performance of \tech{} under different values of the difficulty smoothing factor $\lambda$ in terms of Pass@1 and AvgPassRatio metrics.
First, we observe that, as $\lambda$ increases, performance on both metrics first improves and then degrades. 
The best performance is achieved at $\lambda = 0.6$, where \tech{} reaches 66.50\% and 77.81\% in terms of Pass@1 and AvgPassRatio, respectively. 
This suggests that a training mixture composed of 60\% difficulty-based sampling and 40\% random sampling strikes an effective balance between focusing on difficult samples and preserving previously acquired knowledge, thereby mitigating catastrophic forgetting.

Notably, when $\lambda = 0$, the difficulty-based sampling mechanism is completely disabled and \tech{} reduces to a variant that relies solely on random sampling combined with our requirement optimization component. 
Even in this setting, \tech{}$_{\lambda=0}$ still outperforms the RSCRL baseline by 0.87\% and 0.81\% in terms of Pass@1 and AvgPassRatio, respectively. 
This finding is consistent with RQ2 and further validates the effectiveness of our requirement optimization component.

In contrast, when $\lambda = 1.0$, \tech{} attains only 64.30\% Pass@1 and 75.59\% AvgPassRatio values, representing a moderate drop compared to the optimal $\lambda = 0.6$.
This indicates that relying exclusively on difficulty-based sampling causes the model to lose exposure to previously seen samples, leading to severe knowledge forgetting and degraded generalization.
Due to computational constraints, we evaluate $\lambda$ using a step size of 0.2 over six settings in $[0, 1.0]$. 
A finer-grained exploration of this hyper-parameter will be considered in future work.

\section{Discussion}
\label{sec:discussion}

\subsection{Repeated Experiments}

\begin{table}[htbp]
  \centering
  \tabcolsep=2.2mm
  \caption{Repeated experimental results of \tech{} on Qwen2.5-Coder-3B in term of Pass@1 ($\uparrow$).}
  \label{tab:repeate_exp}
    \begin{tabular}{ccccccc}
    \toprule
    \textbf{Repeat} & \textbf{HumanEval} & \textbf{HumanEval+} & \textbf{MBPP}  & \textbf{MBPP+} & \textbf{LiveCodeBench} & \textbf{Avg.} \\
    \midrule
    Repeat-1  & 82.93\% & 77.44\% & 79.10\% & 67.46\% & 25.60\% & 66.51\% \\
    Repeat-2  & 82.32\% & 76.83\% & 78.31\% & 67.20\% & 25.50\% & 66.03\% \\
    Repeat-3  & 82.93\% & 76.22\% & 78.31\% & 66.67\% & 25.02\% & 65.83\% \\
    Repeat-4  & 83.54\% & 77.44\% & 78.31\% & 66.93\% & 25.59\% & 66.36\% \\
    Repeat-5  & 82.93\% & 78.05\% & 78.57\% & 66.14\% & 25.50\% & 66.24\% \\
    \bottomrule
    \end{tabular}%
  \label{tab:addlabel}%
\end{table}%

To mitigate the impact of inherent stochasticity in both the training and inference phases of LLMs, we assess the robustness and reproducibility of \tech{} through repeated experiments. 
Given the substantial computational cost of full-scale repetition, we independently repeat the RQ1 experiment for \tech{} five times using Qwen2.5-Coder-3B as the base model, while keeping all experimental settings strictly identical across runs.
Table~\ref{tab:repeate_exp} reports the Pass@1 results of these repeated experiments. 
The mean standard deviation across the five runs is only 0.0026, indicating a high degree of stability. 
Furthermore, a \textit{Wilcoxon Signed-Rank Test}~\cite{wilcoxon1963critical} at a significance level of 0.05 yields p-values exceeding 0.0625 for all pairwise comparisons, suggesting that the observed differences across runs are not statistically significant. 
These results demonstrate that the performance gains achieved by \tech{} are stable and reproducible, and are not attributable to favorable random seeds or stochastic effects in the optimization process. 
Consequently, our findings provide strong evidence for the robustness of \tech{} under repeated experimental conditions.


\subsection{Future Work}
\label{subsec:future_work}

Although we have demonstrated the effectiveness of \tech{} through extensive empirical evaluation, several promising directions remain for future investigation.

\begin{itemize}[leftmargin=10pt]
    \item \textbf{Improving Efficiency:} 
    The requirement difficulty perception component relies on executing test cases, which can be time-consuming when the number of test cases is larger. 
    Future work could incorporate test case selection or prioritization strategies to reduce the number of executed tests, as well as parallelized test execution mechanisms to further lower time overhead.

    \item \textbf{Expanding Requirement Attributes:} 
    The current requirement difficulty optimization component is based on five structured requirement attributes. 
    Future work could extend this design by incorporating additional attributes (e.g., code style constraints, code complexity hints, and resource or performance constraints) to further refine optimized requirements. 
    Such extensions could enhance the versatility and applicability of our approach across a broader range of software engineering tasks and domains.

    \item \textbf{Optimizing Difficulty Smoothing:} 
    The adaptive curriculum sampling component employs a fixed difficulty smoothing factor to balance difficulty-based sampling and random sampling, thereby mitigating catastrophic forgetting and encouraging exposure to challenging requirements. 
    Future work could explore dynamic scheduling strategies for this parameter, for example by adjusting the smoothing factor based on reward variance, convergence behavior, or validation performance during training. 
    This could enable more effective curriculum reinforcement learning and further improve model performance.
\end{itemize}

\section{Threats and Validity}
\label{sec:threats}

~\indent
\textbf{Internal Validity.}
This threat primarily stems from the choice of base models and the hyper-parameter configurations used during RL.
To mitigate these threats, we evaluate \tech{} across five diverse LLMs with varying architectures and parameter scales. 
This diversity helps ensure that the observed performance improvements are attributable to \tech{} itself rather than to idiosyncratic properties of a particular model. 
In addition, we maintain identical hyper-parameter settings across all experimental groups to guarantee a fair and controlled comparison.

\textbf{External Validity.}
This threat mainly arises from the selection of benchmarks. 
Although training is primarily conducted on the APPS+ dataset, its broad coverage of algorithmic patterns and difficulty levels makes it representative of a wide range of programming tasks. 
To further assess generalizability, we evaluate \tech{} on five additional, independent benchmarks. 
The consistent performance gains observed across these diverse benchmarks indicate that \tech{} captures generalizable reasoning capabilities rather than overfitting to a specific benchmark. 
Moreover, while this work focuses on function-level code generation, the three components of \tech{} are modular and task-agnostic, making them readily extensible to repository-level code generation and other software engineering scenarios.

\textbf{Construct Validity.}
This threat primarily concerns the choice of evaluation metrics. 
We adopt the widely-used Pass@1 and AvgPassRatio metrics, which is the standard measure for evaluating functional correctness in code generation. 
Particularly, by focusing on first-attempt generation accuracy, we reduce the influence of stochastic sampling effects and provide a direct assessment of the model's ability to generate the correct code satisfying programming requirements.

\section{Conclusion}
\label{sec:conclusion}
This paper draws inspiration from software requirements engineering and presents \tech{}, a novel requirement-aware curriculum reinforcement learning framework for improving the training effectiveness of LLMs in code generation tasks. 
The proposed framework systematically integrates three key components: requirement difficulty perception, requirement difficulty optimization, and adaptive curriculum sampling. 
Together, these components address fundamental limitations of existing CRL approaches, including misaligned requirement difficulty estimation, underutilization of challenging training instances, and instability caused by abrupt curriculum transitions.
To validate the effectiveness of \tech{}, we conduct comprehensive experiments on five state-of-the-art LLMs across five widely-adopted benchmarks. 
Experimental results demonstrate that \tech{} consistently outperforms five advanced baselines across multiple evaluation metrics, confirming its effectiveness in enhancing LLM-based code generation performance. 
These findings highlight the importance of requirement-aware design in curriculum reinforcement learning and suggest promising directions for applying this paradigm to broader software engineering tasks.

\section{Data Availability}
\label{sec:Data_Available}
We anonymously share a comprehensive replication package to facilitate reproducibility at~\cite{rercl_code}.


\bibliographystyle{ACM-Reference-Format}
\bibliography{references}

\end{document}